\documentstyle[12pt,aaspp4]{article}

\def\RSUN{\rm R_{\odot}}

\begin{document}

\title{The Nature of Solar Polar Rays}
\author{Jing Li, David Jewitt, and Barry LaBonte}
\affil{Institute for Astronomy, 2680 Woodlawn Drive, Honolulu, HI 96822}
\authoremail{jing@ifa.hawaii.edu}

\begin{abstract}
We use time series observations from the SOHO and Yohkoh spacecraft to study solar polar rays. Contrary to our expectations, we find that the rays are associated with active regions on the sun and are not features of the polar coronal holes. They are extended, hot plasma structures formed in the active regions and projected onto the plane of the sky above the polar coronal holes. We present new observations and simple projection models that match long-lived polar ray structures seen in limb synoptic maps. Individual projection patterns last for at least 5 solar rotations. 
\end{abstract}


\section{Introduction}
Polar coronal rays are bright ray-like or arch-like features first seen above the poles of the sun during total solar eclipse. They have been described in the literature for at least a century (e.g. \cite{maunder1901}) and they are often considered as features of the solar polar coronal holes. Several studies have been conducted to determine the physical properties of these high latitude rays (\cite{fisher1995}, \cite{m1996}). While the classical literature tends to use the terms ``ray'' and ``plume'' interchangably, the recent studies show that the polar rays have physical properties distinct from those of polar plumes. For example, the polar rays are hotter than the background polar hole gas ($2.6\times 10^6$ K vs. $0.7\times 10^6$ K; \cite{m1996}) while the plumes ``are cooler than the background by up to (but not more than) 30\%'' (\cite{deforest1997}). The reasons for the differences between polar rays and polar plumes have not yet been closely examined. Here, we use time series observations made with EIT/SOHO and SXT/Yohkoh between 1996 and 1999 to study the morphology and time dependence of polar rays.

\section{Data Analysis}
The observations are based on extreme ultra-violet and soft X-ray full sun disk images taken by the EIT and SXT instruments onboard SOHO and Yohkoh, respectively. About 4 EIT images per day from 1996 through 1998 were used in our study. Each EIT image was processed for flat-fielding and filter-correction. The processed images were used to build limb synoptic maps. The emissions were extracted around the solar limb $3^\circ$ wide in polar angle and $0.03\RSUN$ in altitude. Successive images provide the time domain for this study, leading to the limb synoptic maps. Figure 1 shows a limb synoptic map made from more than 2000 EIT image pairs 195/171 (FeXII/FeIX,X) taken between January 1996 and June 1998. Low temperature areas such as the polar holes appear dark while high temperature areas such as active regions are bright in Fig.1. To clearly display both southern and northern polar holes the range of polar angles (PA) is extended by $90^\circ$ from $0^\circ - 360^\circ$ to $0^\circ - 450^\circ$ in Figure 1. We used SXT/Yohkoh data in movie form to confirm the distribution of the hot plasma associated with active regions moving across the solar disk. 

\begin{figure}[p]
\plotfiddle{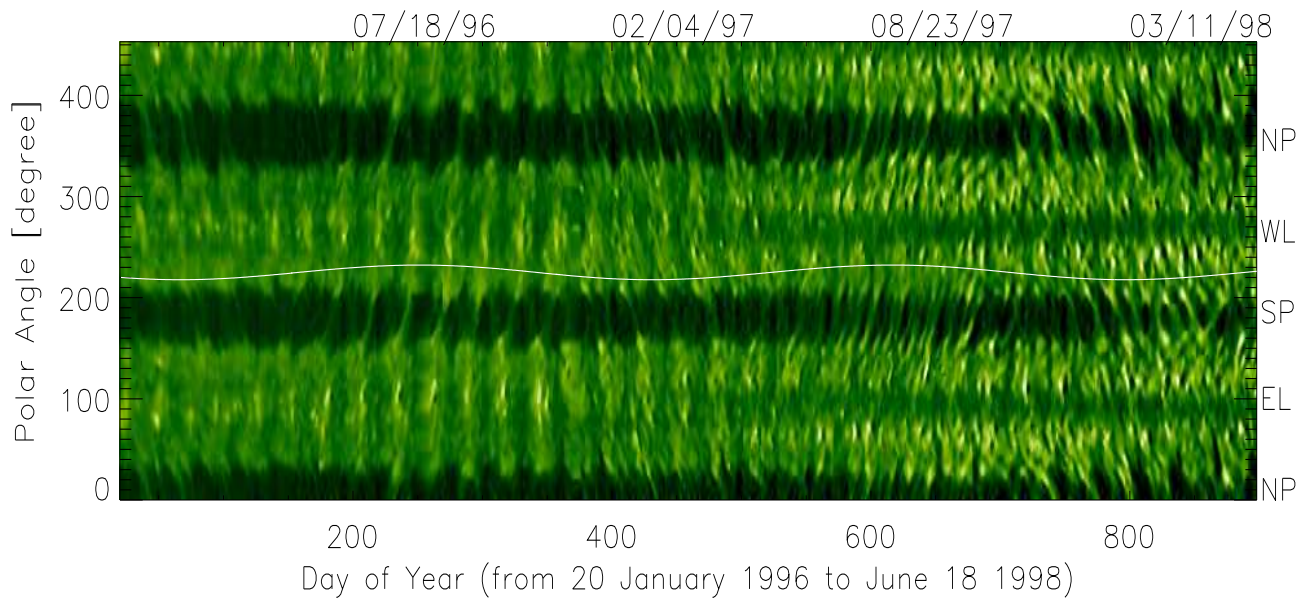}{12cm}{0}{100}{120}{-300}{-400}
\caption{Coronal limb synoptic map at 1.0 $\RSUN$ from EIT 195/171 (FeXII/FeIX,X) images in the period January 1996 to June 1998. The Day of Year (DOY) is indicated along the lower x-axis. The dates are marked along the upper x-axis. The polar angle (PA) is on the y-axis. Zero limb angle corresponds to the north pole (NP), while the east limb is at $90^\circ$ (EL). To display the entire southern ($180^\circ$) and northern ($0^\circ$ and $360^\circ$) polar holes, the polar angle is extended by $90^\circ$. The heliographic latitude is plotted with a solid curve.}
\end{figure}

\section{Results}
Figure 1 shows a number of noteworthy features. The coronal polar holes appear as dark bands between $150^\circ \leq$ PA $\leq 210^\circ$ and $320^\circ \leq$ PA $\leq 390^\circ$, extending from day of year DOY 0 to about DOY 450 (c.f. \cite{li2000}). For DOY $> 450$, the polar holes are filled in by a network of multiply crossing rays. Brighter equatorial gas fills the remaining PAs. An active region (PA $\sim 100^\circ$ on the east limb and $\sim 260^\circ$ on the west limb) persists over 13 full rotations from DOY$=0$ to DOY$=360$. Fingers of emission encroach from low latitudes on the polar hole boundary. After DOY$=210$, these fingers bridge the polar holes, forming a sinusoidal feature that persists for at least 5 rotations. This is dramatically longer than the $\sim 20$ hour lifetime reported for polar plumes (\cite{llebaria1998}). Sinusoids appear simultaneously in the North and South polar holes. At any given time after DOY$\sim$450, the polar holes are crossed by several overlapping sinusoids making a study of the morphology more difficult. In Fig. 2 we enlarge a portion of the limb synoptic map from $210\leq$ DOY $\leq350$ days through EIT 284 (FeXV) in which the ray morphology is particularly clear. The polar ray links successive appearances of the active regions on the East and West limbs, and clearly draws a connection between the polar rays and the active regions.

\begin{figure}[p]
\plotfiddle{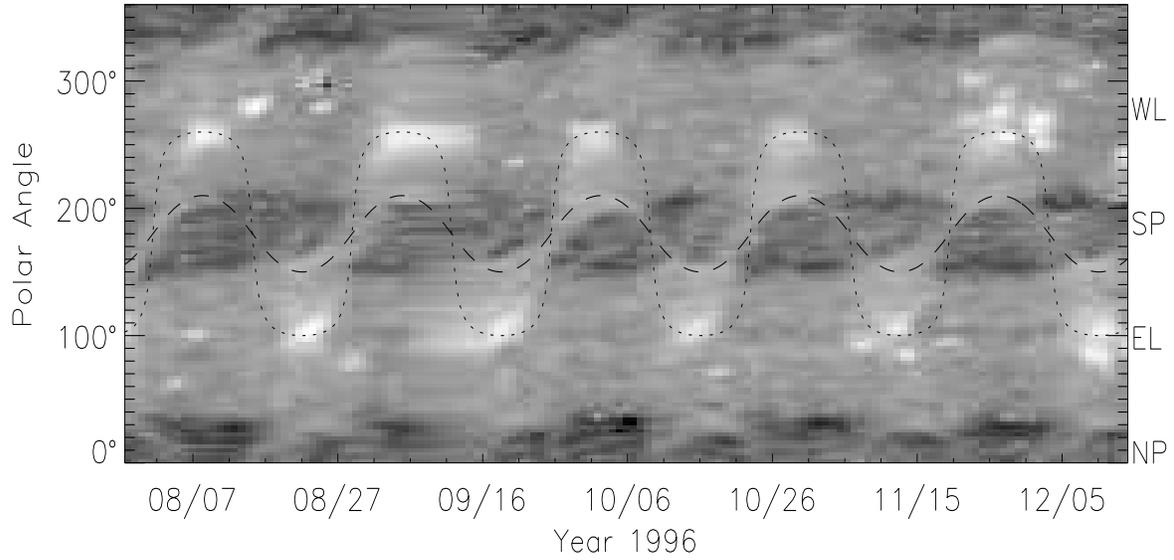}{10cm}{0}{100}{120}{-300}{-400}
\caption{A portion of the synoptic chart from EIT 284 \AA~from DOY$=210$ to DOY=$350$, corresponding to July to December 1996. The dashed curve shows Eq. (1) with $\theta_0=60^\circ$, the dotted curve corresponds to $\theta_0=10^\circ$. The notations on the right side of the figure have the same meaning as those in Fig.1. NP is Northern Pole, EL is East Limb.}
\end{figure}

To model the polar ray sinusoids, we consider the apparent sky-plane motion of a radial coronal structure as it is carried round by the solar rotation. The position angle of such a structure is given by
\begin{equation}
\psi~=~\tan^{-1}\left[\frac{\tan (\theta_0)}{\cos (\alpha)}\right]
\end{equation}
where $\theta_0$ is the heliocentric latitude and $\alpha$ is the longitude of the feature measured from the east limb,
\begin{equation}
\alpha~=~\frac{2\pi}{P(\theta_0)}(t_0+t)
\end{equation}
Here, $P(\theta_0)$ is the coronal rotation period at latitude $\theta_0$, $t$ is time and $t_0$ is the time when the feature crosses the east limb. In Eq. (1) we have neglected the effect of the tilt of the solar equator relative to the ecliptic plane. By inspection, we take $P(\theta_0)=27.5$ days (compatible with independent measurements by \cite{inhester1999}). Eq. (1) is plotted on Fig. 2 for two sample values $\theta_0=10^\circ$ (dotted curve) and $60^\circ$ (dashed curve). The $\theta_0=60^\circ$ curve resembles the polar sinusoid rather closely, and suggests that this feature is due to a nearly radial coronal ray rooted outside the polar hole but carried across it (in projection) by the solar rotation. Models with $\theta_0>60^\circ$, the latitude of the boundary of the polar coronal hole at this epoch, do not fit the shapes of the polar sinusoids. The $\theta_0=10^\circ$ curve fits the polar sinusoid less well but encompasses a broad swath of near-equatorial emission that is associated with the active regions. Together, the two curves strongly suggest that the polar rays are merely projections of more equatorial gas above the limb of the sun at high latitudes. 

A more detailed but still simplistic 3-dimensional corona model was used to further test this possibility. Examination of individual images from EIT and SXT suggested the following representation. We take the active region to be a source of hot gas, which occupies a thick sheet of longitudinal extent $\bigtriangleup \alpha$. The boundaries of the region in the meridional plane were assumed to be parabolic with a vertex at the active region. The radial variation of the electron density is assumed to follow
\begin{equation}
n_e(r)~=~n_{e0}(\RSUN)\exp\left[ -\frac{1.5\times10^7}{T_e}\left( 1-\frac{\RSUN}{r}\right) \right]
\end{equation}
which is consistent with hydrostatic equilibrium. We take $T_e=2.5\times 10^6$ [K] and $n_{e0}(\RSUN)=3\times 10^9$ [$cm^{-3}$]. The assumption of hydrostatic equilibrium is not critical to our result; any extended distribution of gas that projects above the pole in the plane of the sky gives the same qualitative features. The coronal model outside the active region is taken from our earlier work (\cite{li2000}) and consists of separate parts for the equatorial regions and polar holes. We integrated along the line of sight (including the effects of the tilt of the solar rotation axis) to construct model images of the corona as functions of the solar rotation angle. Figure 3 shows a sample model image in which $\alpha = 30^\circ$. Finally, we constructed limb synoptic maps from the models in the same way as for the data. Examples are shown in Figure 4.

\begin{figure}[p]
\plotfiddle{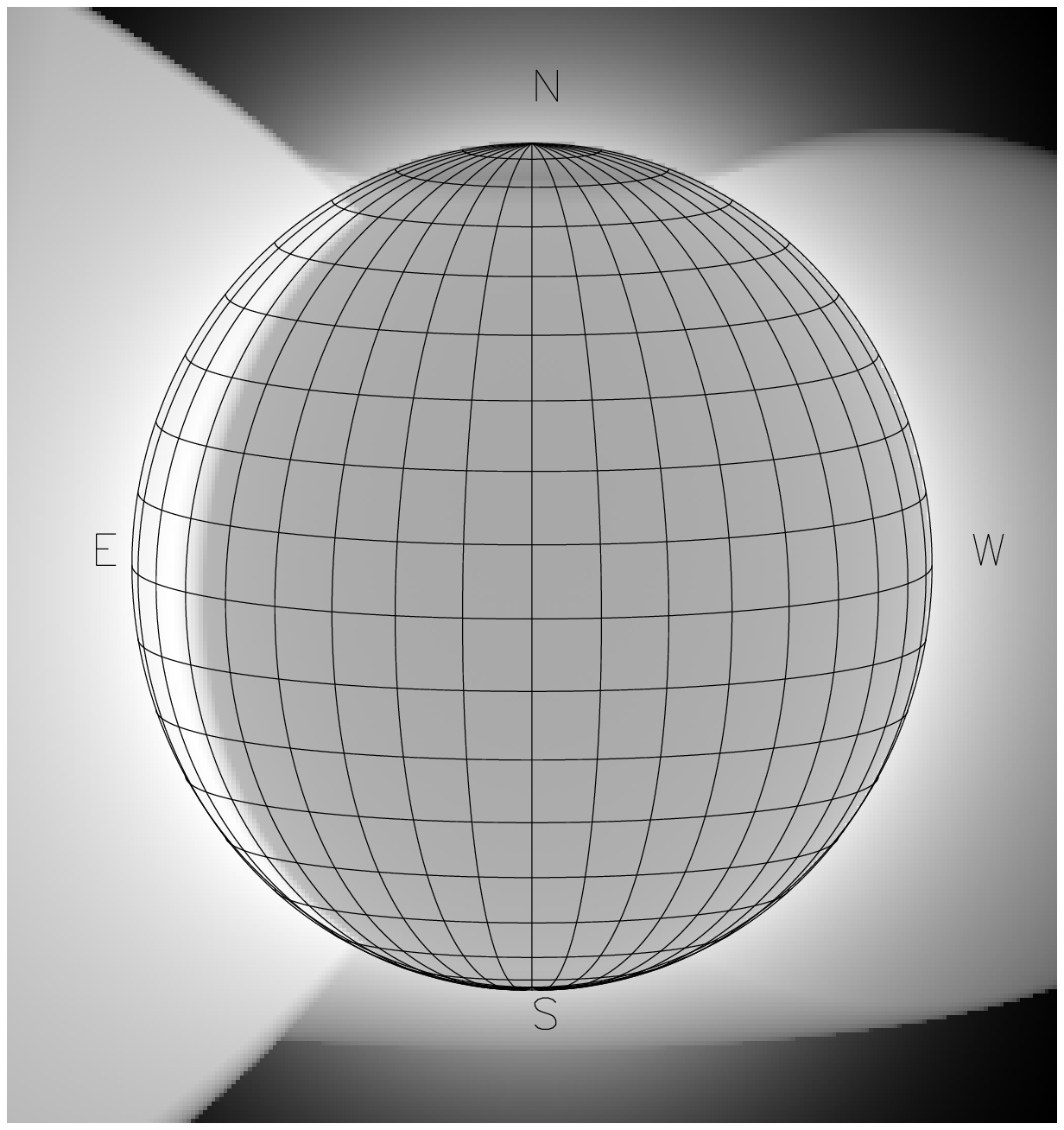}{10cm}{0}{100}{100}{-300}{-400}
\caption{Sample model image in which the active region is located at $30^\circ$ longitude and the north pole of the sun is tilted towards Earth by $7.^\circ 25$ (i.e. $B_0=+7.^\circ 25$). Image has north to the top, East to the left, and the grid spacing is $10^\circ$.}
\end{figure}

\begin{figure}[p]
\plotfiddle{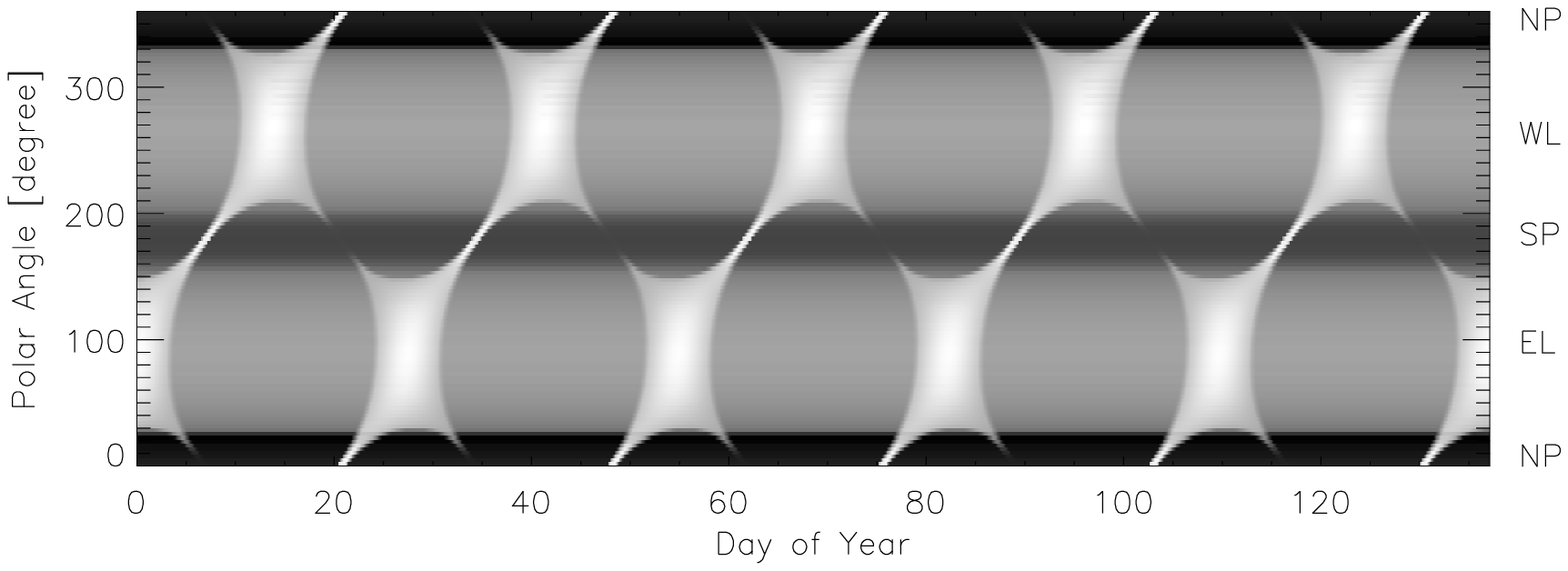}{8cm}{0}{100}{120}{-300}{-400}
\plotfiddle{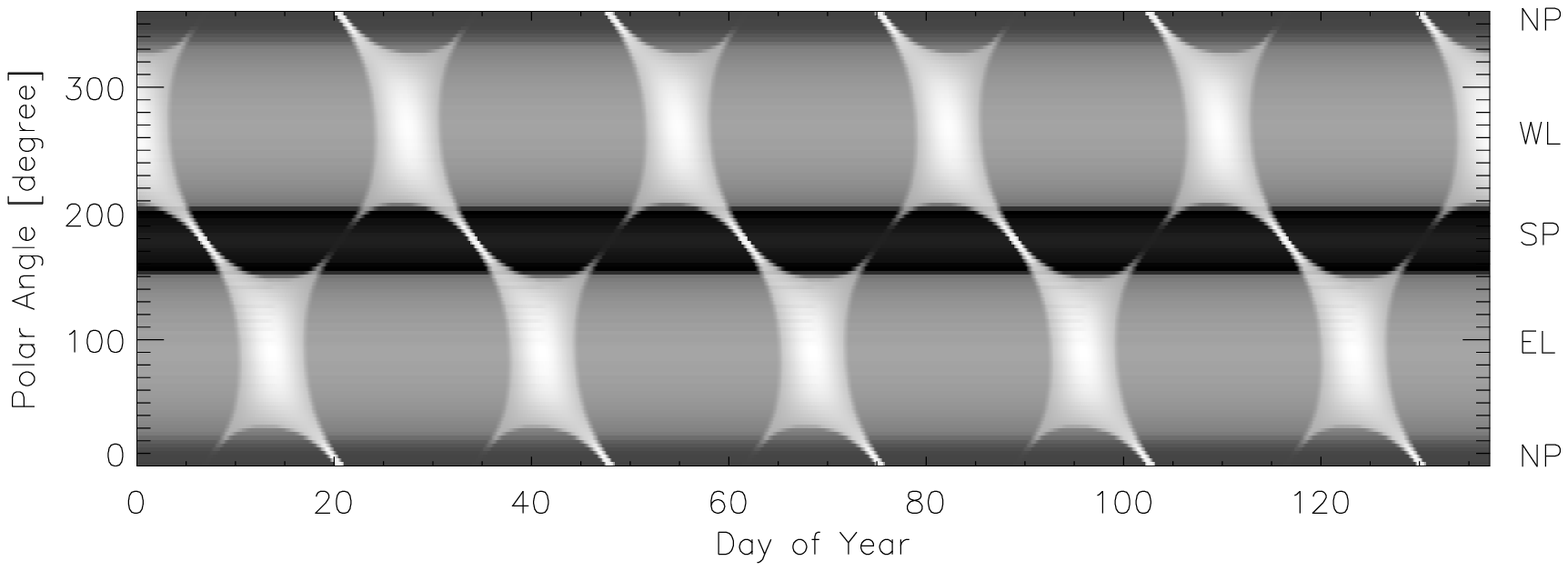}{8cm}{0}{100}{120}{-300}{-350}
\caption{Simulated limb synoptic map made from the model of Figure 3. A rotational resolution of $1^\circ$ was used. About 5 solar rotations are included here. The panels show a) $B_0=+7.^\circ 25$ and b) $B_0=-7.^\circ 25$. Note the reversal in the sense of the asymmetry of the brightness of polar rays.}
\end{figure}

In general, the 3-dimensional models behave exactly as expected from simple geometrical considerations (Eqs. 1 and 2). When the heliographic latitude $B_0=0$, gas from an active region is visible in projection above the polar holes provided it reaches a radial distance
\begin{equation}
r~=~\frac{\RSUN}{\sqrt{\cos^2\theta_0\cos^2\alpha+\sin^2\theta_0}}
\end{equation}
where $\theta_0$ is latitude and $\alpha$ is longitude as before. A plasma structure on the meridian ($\alpha=90^\circ$) at $\theta_0=10^\circ$ must extend to $r\sim 5.8~\RSUN$ in order to project above the pole, while a feature at $\theta_0=60^\circ$ needs to rise only to $r\sim 1.15~\RSUN$ in order to be seen. Therefore, the polar rays are most likely the high latitude edges of extended equatorial structures seen in projection. We used EIT 171 \AA~(Fe IX,X), 195 \AA~(Fe XII) and 284 \AA~(Fe XV) images to estimate the temperatures of the polar rays (e.g. Fig. 1). We find that the rays are hotter than the background polar hole gas, consistent with independent results (\cite{m1996}) and with an origin in more nearly equatorial regions. The observation that polar rays appear simultaneously in the North and South holes (see Fig. 1) is naturally explained in this model if the active region is very extended in latitude, consistent with direct images of the active regions when on the solar limb.

Close inspection of the polar sinusoids in Figures 1 and 2 reveals an additional, more subtle long term variation in morphology. The rising (PA increasing with time) and falling (PA decreasing with time) parts of the polar sinusoids have unequal brightness. In the interval $210\leq DOY \leq 350$, the rising branch is systematically brighter than the falling branch, for both northern and southern features. The sense of this asymmetry reverses in the period $350\leq DOY \leq 530$ (Fig. 1). In fact, Fig.1 shows that the asymmetry alternates in phase with the heliographic latitude ($B_0$), suggesting that projection effects are responsible.

Our 3D coronal model, which includes $B_0$, shows exactly this behavior (see Figure 4). When the north pole is tilted towards the Earth (Figure 4a), the model polar rays are brighter on the rising branch than on the falling branch. The opposite pattern is reproduced in Figure 4b when the north pole is tilted away from the Earth, providing a realistic match to the data. When the north pole is tilted towards the Earth, the hot active region gas must rise higher on the far side of the sun in the southern hemisphere (and on the near side in the north) to meet the line of sight, and so appears fainter than front-side gas on the line of sight at lower altitudes. Therefore, the bright rising segments near DOY $\sim242$ (see Fig. 2, from August 23 to September 4, 1996) identify an active region on the visible hemisphere of the sun, as is seen. The periodically reversing brightness asymmetry is a product of projection effects in equatorial plasma.

As the solar cycle approaches its activity maximum, the low temperature gas in the polar holes is progressively obscured from view by hot equatorial gas projected to high latitudes. This is well seen in Fig.1 where, after DOY $\sim550$, both polar holes are covered by a worm-like network of bright polar rays. The basic morphology of individual rays appears unchanged in the $2\frac{1}{2}$ years of this study: only the number of polar rays increases towards solar maximum. Some previous reports have claimed that the polar holes vary in size and strength through the solar cycle. Our present results suggest instead that the morphology and visibility of the polar holes are largely controlled by the number and strength of active regions far outside the holes and by the associated high altitude gas that projects across the holes as polar rays. In future work, we hope to explore this possibility.

\section{Summary}

1)	Limb synoptic maps of the sun in the period 1996-1998 show long-lived, periodic features in the inner solar corona.

2)	Polar rays appear clearly in such maps with a morphology that shows them to be hot gas structures from active regions projected above the polar holes. Unlike polar plumes, which are clearly rooted in the polar coronal holes, the polar rays are not physically associated with the holes.

3)	The correlated appearance of polar rays above both north and south holes shows that these features are caused by equatorial plasma structures with a latitudinal extent comparable to the solar diameter.

4)	Annually reversing asymmetries in the polar ray brightness on limb synoptic maps are caused by projection effects due to the varying heliographic latitude. 

5) 	Individual polar rays are long lived (more than 5 rotations in this study) and increase in number as the solar activity cycle approaches maximum. The longevity of the polar rays is a reflection of the longevity of active areas that are large enough to be seen in projection above the solar poles. 

6)	Variations in the visibility of the coronal polar holes are largely controlled by projection of hot gas from more equatorial regions. Polar plumes are features best seen at solar minimum. Polar rays are most prominent near solar maximum. 

\acknowledgments
We thank Jeff Kuhn for initially suggesting the data analysis, and Jeff Newmark for his help with the extraction of a large amount EIT data. We thank the referee, Craig DeForest, for his comments. This work was supported by NASA grant 5-4941 and NASA contract NAS 8-40801 for Yohkoh SXT. EIT/SOHO is a joint ESA-NASA program. Yohkoh is a mission of ISAS in Japan.

\end{document}